\documentclass
[pre,aps,twocolumn,superscriptaddress,amsmath,showpacs,floatfix]{revtex4}
\usepackage{graphicx}
\usepackage{epsf}
\def\ignore#1{}
\begin{document}
\title{Unusual percolation in simple small-world networks}
\author{Reuven Cohen}
\affiliation{Department of Mathematics, Bar-Ilan University, Ramat-Gan 52900, Israel}
\author{Daryush Jonathan Dawid}
\affiliation{Brick Court Chambers, Essex Street, London WC2R 3LD, UK}
\author{Mehran Kardar}
\affiliation{Department of Physics, Massachusetts Institute of
Technology, Cambridge, Massachusetts 02139}
\author{Yaneer Bar-Yam}
\affiliation{New England Complex Systems Institute, Cambridge, Massachusetts 02138}
\date{\today}
\begin{abstract}
We present an exact solution of percolation in a generalized
class of Watts--Strogatz graphs defined on a 1-dimensional
underlying lattice. We find a non-classical critical point in
the limit of the number of long-range bonds in the system going to
zero, with a discontinuity in the percolation probability and a
divergence in the mean finite-cluster size. We show that the
critical behavior falls into one of three regimes depending on the
proportion of occupied long-range to unoccupied
nearest-neighbor bonds, with each regime being characterized
by different critical exponents. The three
regimes can be united by a single scaling function around the
critical point.
These results can be used to identify the number of long-range links
necessary to secure connectivity in a communication or transportation chain.
As an example, we can resolve the communication problem in a 
game of ``telephone''. 
\end{abstract}
\widetext
\maketitle

\section{Introduction}
\subsection{Small-world networks}
In a 1967 study on social networks~\cite{Milgram}, Stanley Milgram
found that on average a randomly chosen person in the midwest USA
could be connected to a target person in Massachusetts through a
string of first-name acquaintances in only six steps. While this
notion of ``six degrees of separation'' rapidly acquired folkloric
status in popular culture~\cite{facebook}, 
the structure of the underlying networks
remained largely unexplored until recently. The ``smallness'' was
attributed to the logarithmic scaling with graph size
of distances between nodes on random graphs~\cite{bollobas}.
Actual social networks are far from random, however. Today,
social and other networks are frequently described as consisting of
mutually-interconnected local groups together with some far-flung
ties. Interest in the properties of such
partially-ordered networks was sparked by the work of Watts and
Strogatz~\cite{WS}, who in 1998 introduced and studied numerically
a ``small-world" network model in which a controlled degree of
disorder is introduced into initially ordered graphs by randomly
rewiring some fraction of their links. A schematic representation
of a small-world graph is depicted in Fig.~\ref{example}b.

Watts and Strogatz's observation that even a small degree of randomness
changes the scaling of the minimum graph distance between
randomly-chosen nodes, $l$, with the total number of nodes, $N$,
from the linear behavior of ordered graphs to the logarithmic
scaling associated with random graphs, provoked a flurry of
interest. The scale dependence of $l$, has been obtained by
renormalization-group~\cite{RG} and mean-field analysis~\cite{MFS}.
In particular, Moukarzel~\cite{ICL}
showed that for systems of size $L$ on a $d$-dimensional underlying
lattice there is a crossover length $r_c\sim\log(p L)$,
where $p$ is the density of shortcuts, such that on a scale $r<r_c$,
$l(r)$ scales as $r$, whereas on larger scales $l(r) \sim r_c$.

Today, the term ``small-world" has come to describe any system
displaying a combination of strong local clustering with a small
graph diameter.  Mathias and Gopal~\cite{howwhy} have shown that
optimization of a regular graph for high connectivity and low
total bond length gives rise to small-world behavior.
Interestingly, the optimized graphs are distinguished by the
generation of a relatively small number of highly-connected ``hub"
vertices, rather than a random distribution of long-range links as in the
Watts--Strogatz model. Hub-dominated networks are also generated by
a model for a dynamically expanding graph proposed by Albert and
Barab\'asi~\cite{barabasi} to explain the scale-free (i.e. power-law)
distribution in vertex connectivity common to many real-world
networks, such as the World-Wide Web and the science
citation database.
For scale free networks,
it has been shown~\cite{BR02,cohen,mendes}, that the network
is even smaller than small world networks, i.e., has a sub-logarithmic
average distance between nodes.
A further study of real-world networks by Amaral {\it et al.}~\cite{classes} discerned
two other classes of behavior in small-world networks in addition
to scale-free graphs, characterized by truncated power-law and
rapidly-decaying connectivities respectively, and proposed how
such behavior can arise due to constraints placed on the
Albert-Barab\'asi model.

In addition to work on the properties of small-world networks
themselves, recent attention has focused on the behavior of
physical systems defined on small-world graphs, including Ising
models~\cite{BW}, neural networks~\cite{nn}, and random
walks~\cite{rw1,rw2}. There has also been work on models
of disease epidemics using site and bond percolation defined on
small-world graphs (See, e.g., Refs.~\cite{MN1,KA01,KE05}) as described below.
Evolving contagion processes on small-world networks have
been investigated in Ref.~\cite{rauch}, and the dynamical response
of complex networks due to extenal perturbations is
studied in Refs.~\cite{stimuli,dynamics}. The analysis of
implications of small-world networks for goal directed
social network behavior is presented in Ref.~\cite{manage}.
For some contemporary reviews of small-world networks
and networks in general see, e.g.,
Refs.~\cite{Barabasi_rmp_review,vespignani_book,Mendes_book}.

The wide ranging interest and implications of small world networks
have also led to a need for precise understanding and, where possible,
exact solutions of the fundamental properties of these networks.
In this paper we focus on identifying unusual properties that arise
in the case of small world networks on an underlying one dimensional
lattice. We provide exact solutions for a generalized model and study
the behavior in the limit of few long range links --- a limit we find
is very sensitive to the manner in which it is obtained.
We also point to practical implications for securing connectivity
in one dimensional transportation and communication chains, showing
how failure of some local links is overcome by the existence of long
range links.

\subsection{Percolation on small-world graphs}

Percolation is quite sensitive to the geometry of the underlying lattice.
One dimensional lattices are generally inappropriate for representing the
graph underlying social networks, however, they may be appropriate in
specific contexts. Consider, for instance, a chain of islands,
cities, or nodes in an ad hoc or wired network situated in a nearly
linear formation, say along a road. Such a configuration is extremely
sensitive to breakdown, if even a small number of the links
connecting the nodes is removed. This follows from the well known result
that one dimensional systems only percolate in the limit of
zero failure rate. To fortify the structure of such a network,
random long range links may be added that can backup missing
short range bonds. For road transportation long range bonds
may be added by airlines or boat lines, for wired communications
long range bonds might be provided by microwave or satellite
communication links. The important question: ``How many long range
links are needed to replace an (expected) number of short range
failed links?'' is addressed in this manuscript.

For a simple ring structure it will be shown that for
any constant ratio of long range bonds added to replace
removed short range bonds, some nodes remain isolated
from the rest. However, if only a quarter of the number
of the failed short range bonds exist as long range
shortcuts the chain is not broken---a giant connected
component exists comprising a constant fraction of the nodes.

As an example, this shows how to achieve reliable communication in a game of
``telephone.'' In the game, one person tells a second person something,
the second tells a third, and so on down a chain. For a long chain,
and even a reasonably short chain, errors garble the communication.
Our results show that if there are also long range links of communication
along the chain, comprising a quarter of the short range links, the random
errors that are formed do not cause errors in the answer.

We note the difference in behavior of short range and long
range fortifying links. Short range fortification in a
$k$-ring architecture, in which each node is connected to its
$2k$ nearest neighbors leads to an exponentially decreasing
probability of disconnection in $k$. Still, in the limit of long
enough chains, a fixed $k$ implies that a chain always breaks.
Moreover, this method
of reinforcement is expensive in the number of links needed.
The probability of a node being disconnected from all nodes
in one direction is $p^k$. A lower bound on the probability
of disconnecting the ring at $n$ locations can thus be
approximated by $Np^k/n$. For any constant $p$ and any $n$,
as large as desired, keeping this probability vanishingly
small requires at least $k\sim \ln N$. Thus, resiliency
to random failure requires at least $O(N\ln N)$ bonds
(corresponding to the known collapse of the one dimensional
percolation), whereas, as will be seen, the number
of long range links needed is linear in $N$ (and actually
lower than $N$). It should be noted however, that for random
long range bonds $O(N\ln N)$ bonds are still needed to ensure
that all nodes are connected~\cite{bollobas}.

The best practical solution may be to combine short range
and long range reinforcement.
We find that for $k>1$, and small removal rates of the short
range bonds, the network will remain connected (up to a
statistically insignificant fraction of the nodes) for any
constant ratio of added long-range bonds. Thus, the addition
of relatively few random long-range bonds can ensure
percolation and global connectivity.

\begin{figure}
\includegraphics[height=8cm]{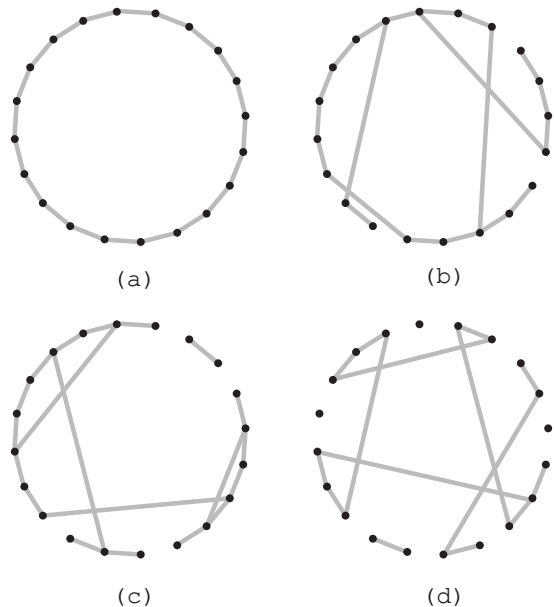}
\caption{\label{example} Examples of small--world graphs: (a) A fully connected
ordered ring with $N=20$ vertices. (b) A graph generated by the
original Watts--Strogatz algorithm with $k=1$ and $p_l=\phi=1/5$,
giving $p_s=4/5$. Note that every long-range link is created by
re-attaching the far end of the broken short-range link.  (c) The
Watts--Strogatz variant used for analytical calculations. Again
$k=1$ and $\phi=1/5$, but this time the addition of long-range
bonds and deletion of short-range bonds is carried out
independently; this makes the appearance of disconnected (finite)
clusters more likely but does not prevent the appearance of
small--world behavior. (d) A generalized Watts--Strogatz graph with
$k=1$, $p=p_l=1/4$, and $n=1/2$, giving $p_s=1/2$.
}
\end{figure}

Recent studies of percolation on small world networks~\cite{sp}
have antecedents in the study of bond-percolation~\cite{KK}
whose methodology is the basis of our analysis.
Small-world-type graphs, with their mixture of ordered local and
random long-range bonds, present the statistical physicist with an
interesting percolation model. It turns out that the construction
of the original Watts--Strogatz (WS) model --- in which shortcuts are created by
reconnecting only one end of an original local bond --- renders it
analytically hard to treat, due to the correlations it introduces
between the distribution of local and long-range bonds. For this
reason, most analytical work is done using a variant of the model
in which long-range bonds are added between randomly chosen pairs of
sites. This change does not affect the small-world behavior of the
resulting graphs, but makes them far easier to treat analytically; we will
refer to both the original and the variant as WS
models. Figure~\ref{example}c gives an example of a variant
WS graph.

Newman and Watts~\cite{sp} have performed both analytical and numerical
analysis of site percolation on such a model with
a local bond coordination number $k$ and a density $\phi$ of
shortcuts added per local bond; no local bonds were removed. Their
results, suggesting site-percolation on WS graphs is similar to
that on random graphs, were confirmed by more rigorous work on the
same system by Moore and Newman~\cite{MN1,MN2}. In addition to
demonstrating that site percolation on small-world networks is
in the same universality class as random graphs (i.e. mean-field),
Moore and Newman presented a formal solution to the bond
percolation problem with independent probabilities of local and
long-range bonds. They were able to solve this for $k=1$ and $k=2$
to show that the percolation transition in such systems also
displays mean-field behavior.

Newman and Moore's results on bond percolation in
small-world systems had in fact been anticipated some 17 years
earlier by Kaufman and Kardar~\cite{KK}, who solved the bond
percolation problem on what would now be regarded as a $k=1$
WS model with general nearest-neighbor and long-range
bond probabilities $(p_s, p_l)$ respectively.
In Section~\ref{review} of
this paper we present a review of Kardar and Kaufman's method and
their results describing the percolation transition. In Section~\ref{main} 
we then use the formalism developed in Ref.~\cite{KK}
to consider the critical percolation behavior of a
standard WS graph by imposing the relation
$p_l=1-p_s$, corresponding to one long-range bond being added to
the system for every local bond removed. This is, as far as we know,
the first analytical treatment of the original Watts--Strogatz model~\cite{WS} 
including rewiring rather than addition of shortcuts.
We then solve the
problem for a generalized graph in which we allow the number of
long-range bonds added for every short-range bond removed to vary
continuously. We also present a brief discussion of percolation on
generalized WS graphs defined on $k$-rings and higher-dimensional
lattices.
An alternative approach to obtaining the results using generating
functions is presented in the Appendix.

\section{Percolation with nearest-neighbor and long-range bonds}\label{review}
\subsection{Model and formalism}
\begin{figure}[htp]
\includegraphics[height=2.5cm]{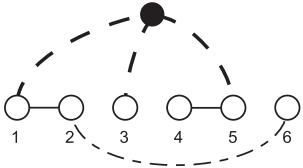}
\caption{\label{cluster}
Example of six sites forming part of a connected cluster
with nearest-neighbor bonds (solid lines), long-range bonds
(dot-dash), and ghost bonds (dashes). The ghost bonds connect sites
(1), (3), and (5) to  the ghost super-site, indicated by
a black circle. For any finite value of $h$, a finite proportion
of sites in the graph will be connected together via ghost bonds
to the super-site, and so all clusters involving ghost bonds must
belong to the infinite cluster. }
\end{figure}

Quite generally, the Kastelyn-Fortuin formalism~\cite{cf} allows us to relate the
bond percolation problem to the $q \rightarrow 1$ limit of the Potts model.
Let us start by considering $q$-state (Potts) spins $s_i$, on a lattice of $N$ sites.
The spins are assumed to be subject to both nearest-neighbor and infinite-range
interactions, with a total energy given by the Hamiltonian
\begin{eqnarray} \label{ham}
-\beta{\cal H}(K,J,h) &=&K\sum_{\langle ij \rangle}\delta_{s_i,s_j}
\nonumber
\\&& +h\sum_i\delta_{s_i,1}
+\frac{J}{2N}\sum_{i,j}\delta_{s_i,s_j}.
\end{eqnarray}
The long-range interaction has to be scaled by $1/N$ to achieve a proper thermodynamic limit,
and  $h$ is a symmetry breaking field.
The cluster-size generating function is given in terms of the
Potts free energy $f(K,h,J)$ by
\begin{eqnarray} \label{Gn}
G(p_s,p_l,h)&\equiv&\sum^\infty_{s=1} n_s (p_s,p_l)
e^{-sh}\\&=&-\left.\frac{\partial f}{\partial q}\right|_{q=1}.
\label{G}\end{eqnarray}
Here the nearest-neighbor
bond occupation probability is $p_s=1-e^{-K}$, the long-range bond
occupation probability is $2p_l/N=1-e^{-J/N}$ (giving an average
of $Np_l$ occupied long-range bonds), $n_s$ is the mean density of
$s$-sized clusters, and $1-e^{-h}$ is the ghost bond probability.
``Ghost bonds'' are the percolation equivalent of a magnetic field
in the Potts model, and can be considered as connecting each site
in the lattice to a single super-site; as a result any non-zero
ghost-bond probability automatically results in the formation of
an infinite (spanning) cluster, since all sites with occupied
ghost bonds form part of the same cluster. An example of a cluster
involving nearest-neighbor, long-range, and ghost bonds is given
in Fig.~\ref{cluster}.

The main quantities of interest in percolation are the percolation probability,
and the mean finite-cluster size:
$P(p_s,p_l)$, is defined as the probability that a site belongs to
the infinite cluster, while $S$ is
the expected size of the cluster of which the site is a member,
averaged over all finite cluster sizes. They are given in terms of
$G(p_s,p_l,h)$ by
\begin{eqnarray}
P(p_s,p_l)&=&1+\frac{\partial}{\partial h}\,G(p_s,p_l, h=0^+),
\label{P_def} \\ S(p_s,p_l) &=&\frac{\partial^2}{\partial
h^2}\,G(p_s,p_l, h=0^+). \label{S}
\end{eqnarray}
To find the free energy of the Potts model, we introduce the
partition function
\begin{eqnarray}\label{partition}
&&Z=\int^\infty_{-\infty}\prod^q_{\alpha=1}
(dx_\alpha\,e^{-(NJ/2)x^2_\alpha})\sum_{\{s_i\}} \\
&&\exp\left(
K\sum_{\langle ij \rangle}\delta_{s_i,s_j}+\sum_i
[(h+Jx_1)\delta_{s_i,1}+\cdots+Jx_q\delta_{s_i,q}]\right).\nonumber
\end{eqnarray}
Carrying out the integrals over ${x}$ and dropping terms
of order $\exp[\log N/N]$, we obtain the partition function for
the original Hamiltonian introduced in  Eq.~(\ref{ham}) as
\begin{equation} \label{part2}
Z=\sum_{\{s_i\}}e^{-{\cal H}(K,J,h)/kT}=e^{-Nf(K,J,h)},
\end{equation}
where $f(K,J,h)$ is the free energy of our Potts model. If instead
we first sum over spins in the partition function of the integrand in Eq.~(\ref{partition}), we obtain
\begin{eqnarray} \label{part3}
Z&=\int^\infty_{-\infty}\prod^q_{\alpha=1} dx_\alpha &\exp[-(NJ/2)\sum_{\alpha=1}^q x_i^2
\nonumber \\ && -Nf_0(K, h+Jx_1,\ldots,Jx_q) ],
\end{eqnarray}
where $f_0(K,h+Jx_1,\cdots,Jx_q)$ is the free energy of the
nearest-neighbor Potts model, ${\cal H}(K, J =0,h_1,\cdots,,h_q)$,
in magnetic fields $h_1=h+Jx_1,\cdots,h_q=Jx_q$.

In the thermodynamic limit $N\rightarrow\infty$, a saddle-point
method~\cite{kardar} relates the two free energies in Eq.~(\ref{part2})
and Eq.~(\ref{part3}) by
\begin{eqnarray} \label{freeenergy}
f&=&\min\bigg[\frac{J}{2}\sum_{\alpha=1}^q x_\alpha^2 + f_0(K,h+Jx_1,\cdots,Jx_q)\bigg]_{\{x_n\}} \\
&=&\frac{-J}{2q}+\min\bigg[Jm+\frac{Jq(q-1)}{2}m^2
+f_0(K,h+qJm)\bigg]_m,\nonumber
\end{eqnarray}
where in the last line we have used the observation that the
$x_\alpha$ are the magnetizations of the Potts
model~\cite{KK,kardar} and introduced a parametrization
\[
x_1=1/q+(q-1)m,\quad x_2=\cdots=x_q=1/q-m. \]

Setting $q=1$ and using Eq.~(\ref{G}), we obtain the cluster-size
generating function for percolation with nearest-neighbor and
long-range bonds,
\begin{equation}\label{KK10}
G(p_s,p_l,h)=-\min[p_l(m-1)^2-G_0(p_s,h+2p_lm)]_m,
\end{equation}
where $G_0(p,h)$ is the cluster-size generating function for
percolation with only nearest-neighbor bonds. We note that
$\bar{m}$, the value of $m$ that minimizes the expression in Eq.~(\ref{KK10}),
is  the percolation probability $P$.

It is simple to calculate the function $G_0(p,h)$ in one dimension
using Eq.~(\ref{Gn}); to form an $s$-size cluster requires a
sequence of $(s-1)$ occupied bonds with an empty bond at each end
to terminate the sequence. Thus $n_s=(1-p_s)^2p_s^{s-1}$, and
performing the sum we find
\begin{equation}
\label{G_0}
G_0(p_s,h)=\frac{(1-p_s)^2}{e^h-p_s}.
\end{equation}

Substituting this result into Eq.~(\ref{KK10}) we obtain the
generating function for bond-percolation on a graph with
nearest-neighbor and long-range bonds,
\begin{equation} \label{KK15}
G(p_s,p_l,h)=-\min\left[
p_l(m-1)^2-\frac{(1-p_s)^2}{e^{h+2p_lm}-p_s}\right]_m.
\end{equation}
This is the result that we shall be using for the rest of
this paper; we note that for percolation in the context of
small--world networks only the $h\rightarrow 0^+$ limit is
relevant.

\subsection{Mean-field percolation}
Given that in the vicinity of the percolation transition we expect
$P=\bar{m}$ to be small, we can perform an expansion of 
Eq.~(\ref{KK15}) in powers of $m$ to give (taking $h=0$)
\begin{eqnarray} \label{KK16}
G(p_s,p_l,0)=&&1-p_s-p_l \min \bigg\{ p_l m^2
\left[1-2p_l\frac{1+p_s}{1-p_s}\right]\nonumber\\
&&+\frac{4p_l^3m^3}{3}
\frac{1+4p_s+p_s^2}{(1-p_s)^2}+{\cal
O}(m^4) \bigg\}_m .
\end{eqnarray}
The phase transition thus occurs when the coefficient of $m^2$
changes sign, giving a transition boundary 
\begin{equation}
2p_l=\frac{1-p_s}{1+p_s}.
\end{equation}
Note that since $m$ cannot be negative, the presence of a cubic
term does not imply a first order transition. Defining
\begin{equation} \label{t}
t=2p_l\left(\frac{1+p_s}{1-p_s}\right)-1,
\end{equation}
we find the critical behavior at the transition to be
\begin{eqnarray} \label{KK18i}
G^{\text{sing}}
&\sim& \left\{
\begin{array}{ll}
0, & t<0; \\ t^3, & t>0;\end{array}\right. \\
\label{KK18ii} P=&\bar{m} \sim& \left\{
\begin{array}{ll} 0, & t<0; \\ t, & t>0;\end{array}\right. \text{ and} \\
\label{KK18iii} S&\sim&t^{-1}.
\end{eqnarray}
Using the standard definitions~\cite{SA}, $G^{\text{sing}}\sim
t^{2-\alpha}$, $P \sim t^\beta$, and $S \sim t^{-\gamma}$, this
gives critical exponents
\begin{equation} \alpha=-1,\quad  \beta=1, \quad  \text{ and } \gamma=1,\label{meanexp}
\end{equation}
typical of mean-field behavior~\cite{kk6,kk7}, in agreement
with the results of Moore and Newman~\cite{MN2,note1}.

\section{Generalized Watts--Strogatz graphs} \label{main}
\begin{figure}[htp]
\includegraphics[height=6cm]{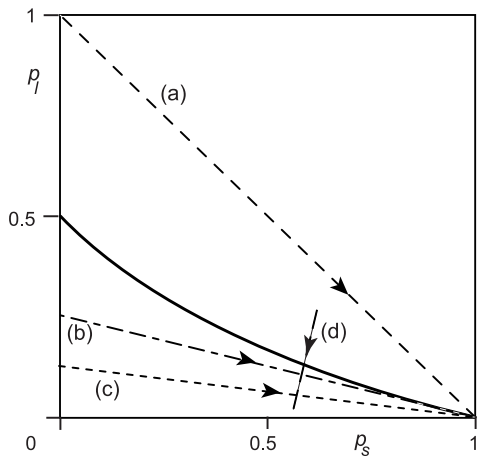}
\caption{\label{traj} Small--world trajectories: The lines (a), (b), and (c)
show examples of small--world trajectories superimposed on the
phase diagram of the more general percolation system. The arrows
show the direction of {\it decreasing} $p$. Trajectory(d) shows a
non-small--world trajectory for which mean-field behavior holds at the transition.
Trajectory (a), with $n=1$, corresponds to the standard
Watts--Strogatz graph and is clearly always percolating.
Trajectories (b) and (c), with $n=1/4$ and $n=1/8$ respectively,
are always in the non-percolating regime.
}
\end{figure}
\subsection{Percolation} \label{N=1}

The results of Section~\ref{review}, describing percolation with
general nearest-neighbor and long-range bond-occupation
probabilities, give a solution to the bond-percolation problem as
defined on a small-world graph~\cite{MN1,MN2}, but are not quite
appropriate to describing the percolative behavior of the graph
itself. Recall that in the Watts--Strogatz model, the proportions
of long-range and nearest-neighbor bonds are not independent, but
rather a fraction $p$ of nearest-neighbor bonds is replaced
with long-range bonds. Thus $p_l$ and $p_s$ are firmly connected
by the relationship $p = p_l=(1-p_s)$. For any
value of $p$, the WS graph occupies a point in
the $(p_s,p_l)$ configuration space of the more general model
which satisfies this relation, and varying $p$ describes a
trajectory through this configuration space (see Fig.~\ref{traj}a).
The appropriate equation for describing the
percolative behavior of the $k=1$ (nearest-neighbor)
Watts--Strogatz graph is therefore given by substituting for $p_s$
and $p_l$  in Eq.~(\ref{KK15}), giving us  the cluster-size
generating function for the WS model as
\begin{equation}
\label{G_WS}
G(n,p,h)=-\min\left[ p(m-1)^2 - \frac{p^2}{e^{h+2pm}-1 +
p}\right]_m.
\end{equation}

\begin{figure}[htp]
\includegraphics[height=14cm]{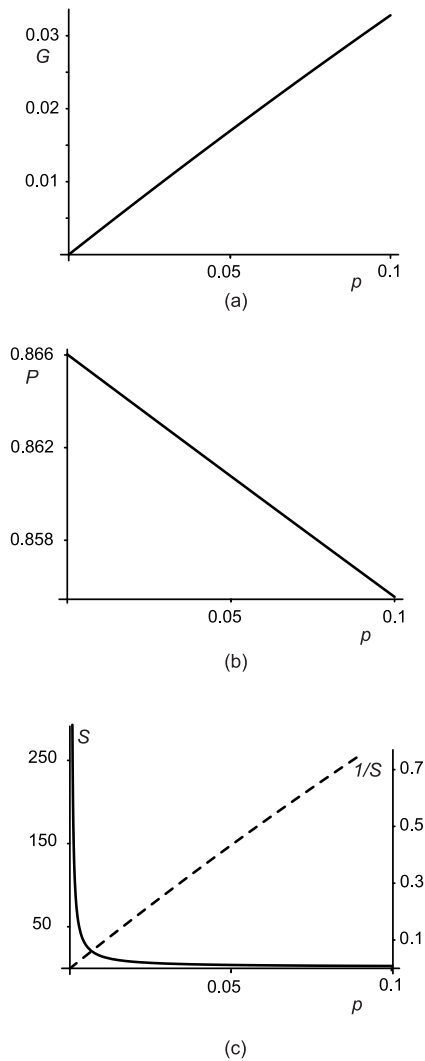}
\caption{\label{numerical}
Numerical results for the standard ($n=1$) Watts--Strogatz
network: (a) Cluster size generating function $G(p)$, (b)
percolation probability $P(p)$, and (c) the mean finite-cluster size
$S(p)$.
}
\end{figure}

According to Eq.~(\ref{t}), the controlling parameter $t$ for this
system is given by $t=3-p \; \Rightarrow \; 2\leq t \leq 3$, and
therefore we expect the system to lie deeply in the percolating
phase for all values of $p$. Nonetheless, it is not difficult to
solve Eq.~(\ref{G_WS}) numerically, and the
resulting graphs of $G$, $P$, and $S$ as a function of $p$ are
illustrated in Fig.~\ref{numerical}. The divergence of $S$ as $p
\rightarrow 0$ indicates that $p=0$ represents some kind of
critical point. Additional evidence for this comes from the fact
that while $P$ comes in linearly to the limiting value
$P=\sqrt{3}/2 \approx 0.866$
as $p\rightarrow 0$, at the exact point $p=0$ the bond
configuration of the system is that of a completely connected
linear chain and therefore $P=1$ is the only physically acceptable
value. Therefore we are faced with a divergence in $S$ and a
discontinuity in $P$ at $p=0$. Taken together with the fact that
$p$ is the only free parameter in the system once we have set
$h=0$, it is reasonable to take $p$ rather than $t$ to be the
appropriate control parameter for percolation
in the WS network.

With this in mind, we can compare the more general results of Eqs.~(\ref{KK18i}), (\ref{KK18ii}) and (\ref{KK18iii})
to the  $p\rightarrow 0$ behavior of the WS network, as indicated by Fig.~\ref{numerical},
\begin{eqnarray}
 G & \sim & p, \nonumber\\
P & \sim & \text{const,} \label{GPS_n=1}\\
S & \sim & p^{-1}. \nonumber
\end{eqnarray}
Evidently,  only $S$ performs according to the mean-field
predictions; $G$ is linear in $p$ rather than the expected cubic
behavior, while $P$ is constant rather than linear. Therefore, in this limit the
WS network exhibits critical exponents $\alpha=1$,
$\beta=0$, and $\gamma=1$ in contrast to the mean-field values
in Eqs.~(\ref{meanexp}). We note for future reference that
the critical exponents found for the WS network
satisfy the scaling exponent identity $\alpha+2\gamma+\beta=2$.

Thus, the above results demonstrate
that the WS graph exhibits an unusual transition as $p \rightarrow 0$.
Rather than the generic mean-field behavior typical of
percolation with general short- and long-range bond
probabilities, the WS network displays a transition
between two percolating phases, with $P$ jumping discontinuously
from a finite value to unity while $S$ diverges continuously as
$1/p$. In the following section, by considering a generalization of
this model we provide an analytical explanation for
this behavior.

\subsection{Scaling} \label{general}
The original choice of Watts and Strogatz, to fix the number of
added long-range bonds to equal the number of short-range bonds
removed, was motivated by the desire to compare the properties of
graphs having the same total number of bonds for different values
of $p$. However, when applying their model to physical systems
there is no reason {\it a priori} to assume this to be the case.
In some cases long-range bonds are more ``costly'' than short-range
bonds and we would expect the relative number of long-range
bonds in the system to be fewer. Alternatively, in the context of
neural connections, suitable training of a growing cortex can
result in a relative increase in the number of long-range bonds.

With these considerations in mind, we now define a generalized
WS graph by the simple expedient of allowing the
number of long-range bonds added for every short-range bond
removed to vary continuously; Fig.~\ref{example}d gives an example
of such a generalized graph. Thus for such graphs, the
relationship between $p_s$ and $p=p_l$ is given by
\begin{equation} \label{p}
p=p_l=n(1-p_s).
\end{equation}
An illustration of some of the trajectories for different values
of $n$ is given in Fig.~\ref{traj}. Substituting Eq.~(\ref{p})
into Eq.~(\ref{KK15}) gives us the cluster-size generating
function for nearest-neighbor generalized Watts--Strogatz graphs,
\begin{equation} \label{Gg}
G(n,p,h)= -\min[f(m,n,p,h)]_m,
\end{equation}
where for convenience we have defined
\begin{equation} \label{f_def}
f(m,n,p,h)= p(m-1)^2 - \frac{(p/n)^2}{e^{h+2pm}-1 + p/n}.
\end{equation}

We now restrict our attention to trajectories along which $P$ is
small in the $p \rightarrow 0$ limit, which we expect to be those
in the immediate vicinity of,  and beneath, the transition line in
Fig.~\ref{traj}. While the restriction on $m$ being small means
that any results derived using it are not strictly applicable to
the $n=1$ case treated numerically in Sec.~\ref{N=1}, we expect
that universal quantities such as critical exponents should agree
in both cases. With this in mind, we make a small-$m$ expansion of
Eq.~(\ref{f_def}) to give the equivalent of Eq.~(\ref{KK16}).

We start with the case $h=0$. The free energy then becomes
\begin{eqnarray} \label{f}
f(m,n,p,h=0)&=&p\left(1-\frac{1}{n}\right)+pm^2(1+2p-4n) \nonumber \\
&&+\frac{4}{3}pm^3(6n^2-6pn+p^2).
\end{eqnarray}
Considering the coefficient of $m^2$, we immediately see that the
percolation transition occurs when
\begin{equation}
n=\frac{1+2p}{4}.
\end{equation}
Therefore in the $p \rightarrow 0$ limit, the transition
line in Fig.~\ref{traj} comes down linearly with a
slope of $-1/4$. As in  Eq.~(\ref{KK16}), the presence of a
cubic term does not imply a first-order transition since
$\bar{m}=P$ is confined to the range $(0,1)$. Defining
$\delta_n=n-1/4$,
and setting $\partial f
/ \partial m=0$ to obtain $\bar{m}$, we find
\begin{equation}
P=\bar{m}=\left\{ \begin{array}{ll} 0 & n<\frac{1}{4}(1+2p)
\\ \frac{4\delta_n-2p}{12n^2-12np+2p^2} & n>\frac{1}{4}(1+2p)
\end{array}. \right.
\end{equation}
This result is only valid for $\bar{m}\ll 1$, and so it follows
that our expression for $f$ in Eq.~(\ref{f}) is only valid for $n
\alt 1/4$.
We can write the singular parts of $G$ and $P$ in the
$p\rightarrow 0$ limit as
\begin{mathletters}
\begin{eqnarray}
\label{Gsing} G^{sing}(p\rightarrow 0;n) & \sim &\left\{
\begin{array}{ll} 0, & \delta_n \leq p/2; \\ p \delta_n^3 , & \delta_n>p/2;
\end{array}\right.
\\ \label{P} P(p\rightarrow 0;n)=\bar{m} & \sim &\left\{
\begin{array}{ll} 0, & \delta_n\leq p/2; \\   4\delta_n-2p, & \delta_n>p/2.
                       \end{array}\right.
\end{eqnarray}
\end{mathletters}

Once again, these results are quite distinct from the mean-field
predictions of  Eqs.~(\ref{KK18i}) and (\ref{KK18ii}). Moreover,
the results for $\delta_n>p/2$, giving  $\alpha=1$ and $\beta=0$
with $P$ tending linearly to a finite constant as $p \rightarrow
0$, are in accordance with the numerically obtained
behavior of Eqs.~(\ref{GPS_n=1}), arguing for the universality of
Eqs.~(\ref{Gsing}) and (\ref{P}) at all $n>1/4$

Note that the result from Eq.~(\ref{P}), that increasing $p$
reduces the percolation probability, should not be
surprising, since the total number of bonds in the system is
proportional to $(p_l+p_s)=1+p(1-1/n)$. Therefore for $n<1$,
increasing $p$ reduces the total number of bonds in the system.
This can also be understood by considering  the convex nature of
the transition line in Fig.~\ref{traj}; all trajectories with
$\frac{1}{4}<n<\frac{1}{2}$ will intersect the line twice, once at
$p=0$ and again at $p\approx 2\delta_n$. Moreover, the second
transition does behave in the mean-field manner, meaning that $P
\rightarrow 0$ linearly as we approach it. Since when $\delta_n
\ll 1$ the distance between the two transitions becomes
vanishingly small, the percolation probability $P$ must decrease
from $P=4\delta_n$ to $P=0$ linearly as we increase $p$ through
the increment from 0 to $\delta_n/2$.

We now turn to calculating the mean finite cluster size, $S$,
which is the equivalent of the susceptibility for a percolation
system.  Equations~(\ref{P_def}) and (\ref{S}) imply that we can
write
\begin{equation} S=\left.\frac{\partial \bar{m}}{\partial h}\right|_{h=0^+}.
\end{equation}
We therefore need to find the $h$-dependence of $\bar{m}$. This is
simple to do if we notice that by defining
\[
\hat{m}=m+\frac{h}{2p},
\]
we may express $f(m,h)$ as
\begin{equation}\label{fh}
f(m,h)=f(\hat{m},0)-\frac{h^2}{4p}+h\hat{m}+h.
\end{equation}
Expanding $G$ for small $p$, $m$, $\delta_n$ and $h$
we find
\begin{eqnarray}
\label{free_en_h}
G=&&p\min\bigg[\frac{3}{4}+\frac{h}{2p}+2m^2(p-2\delta_n) \nonumber\\
&&+\frac{mh}{p}(2p-4\delta_n-1)\nonumber\\
&&+\frac{h^2}{4p^2}(2p-4\delta_n-1)
+\frac{1}{2}\left(m-\frac{h}{2p}\right)^3\bigg]_m.
\end{eqnarray}
Minimizing the free energy with respect to $m$ and expanding
to lowest orders in $p$, $\delta_n$ and $h$, one obtains
the magnetization as
\begin{equation}
\bar m=
\begin{cases}
\frac{h}{p(4p-8\delta_n)},&\delta_n\leq p/2\\
\frac{16\delta_n-8p}{3}-\frac{h}{p(4p-8\delta_n)},&\delta_n> p/2
\end{cases}.
\end{equation}
Substituting this value into Eq.~(\ref{free_en_h}), we get
the free energy expression near the singular point as
\begin{equation}
G=p\left(\frac{3}{4}+\frac{h}{2p}-\frac{3h^2}{8p^2(2\delta_n-p)}\right),
\end{equation}
for $\delta_n\leq p/2$, and
\begin{equation}
G=p\left(\frac{3}{4}+\frac{8^3}{2\cdot3^3}(2\delta_n-p)^3+
\frac{h}{2p}-\frac{3h^2}{8p^2(2\delta_n-p)}\right),
\end{equation}
for $\delta_n>p/2$. This implies that the singular part of the
free energy can be written as
\begin{equation}
G^{\rm sing}=p^4x^3\phi_x(\tilde h),
\end{equation}
where,
\begin{eqnarray}
x&=&\frac{\delta_n}{p}-\frac{1}{2},\\
\tilde h&=&\frac{h}{p^3x^2},
\end{eqnarray}
and $\phi$ is a universal scaling function, obeying,
\begin{equation}
\phi_x(\tilde h)\sim
\begin{cases}
\tilde h^2& x\leq 0,\\
\frac{3}{2}-\tilde h& x>0.
\end{cases}
\end{equation}

Therefore in the  $p \rightarrow 0$ limit, we
find the critical behavior of $S$ as $p \rightarrow 0$ to be
\begin{equation} \label{Scrit}
S (p\rightarrow 0;n) \sim   \left\{ \begin{array} {ll}
   -n/(2\delta_n p) & \delta_n < p/2; \\
   (1/2-n)/(2\delta_n p) & \delta_n  \agt  p/2;\\
   1/(4p^2) & \delta_n=p/2.
   \end{array} \right.
   \end{equation}
The critical exponent for the divergence of $S$ is therefore
$\gamma=1$ for $\delta_n\not =p/2$, and $\gamma=2$ for $\delta_n=p/2$.
These results for $S$ show a number of interesting points. The
first is that for $\delta_n \not = p/2$, $S$ diverges in agreement
both with the mean-field prediction and the numerical result for
$n=1$. We also note that  the amplitude for the divergence at $\delta_n=p/2^+$
is equal to that for $\delta_n=p/2^-$. However,  for
$\delta_n=p/2$ precisely, the percolation transition is accompanied
by a different divergence in $S$. Thus, while $G$ and $P$ display
the same singular behavior for $\delta_n=p/2$ and $\delta_n <p/2$, the
two cases are differentiated by the order of divergence in $S$.

\subsection{Other lattices}
Given the rich behavior of percolation on the simple WS model,
it is  natural  to enquire if similar behavior exists
when the underlying lattice structure is of a higher dimension.
Returning to the considerations of Sec.~\ref{review}, we note
that to treat the higher-dimensional case all we need do is to
substitute the appropriate $G_0$ into Eq.~(\ref{KK10}).
Unfortunately,  the exact form of $G_0$ is not known for
dimensions higher than one (not counting the infinite-dimensional
Bethe lattice), and so an analytic solution of the
higher-dimensional case is not possible by this
method. Nevertheless, enough is known about the properties of
$G_0(p_s,h)$ in higher dimensions to make some general
statements about percolation in such graphs. In particular, close to the
percolation transition we can write a form equivalent to the
expansion in Eq.~(\ref{KK16}), as
\begin{eqnarray}
&&G(p_s,p_l,0)=-p_l+G_0(p_s,0)-\min \nonumber\\
&&\left\{ p_l m^2
[1-2p_lS_0(p_s)]+\frac{4}{3} p_l^3 U_0(p_s)m^3 +
O(m^4)\right\}_m,\nonumber
\end{eqnarray}
where $S_0=\partial^2 G_0/\partial h^2|_{h=0^+}$ is the mean
finite-cluster size of the nearest-neighbor model on the
underlying lattice, and $U_0=-\partial^3 G_0/\partial
h^3|_{h=0}>0$. There will thus be a percolation transition at the
point $2p_l=S_0^{-1}$, again governed by mean-field exponents.

This result provides enough information
to draw some conclusions about the behavior of
generalized WS graphs in higher dimensions. In
particular, in the  one-dimensional case it was the intersection
of small-world trajectories with the critical point at $p=0$ that
gave rise to all the interesting behavior as $p \rightarrow 0$. In
higher dimensions, however, small-world trajectories as defined in
Eq.~(\ref{p}) do not encounter any critical points as $p
\rightarrow 0$. Thus, we would not expect any critical behavior
other than a mean-field transition should they pass
through the transition line (in the $(p_l,p_s)$ space)
at some point other than the origin.

\section{Conclusion and Discussion}
We have examined the percolation properties of a class of
WS graphs generalized by allowing the proportion of
long-range and short-range bonds to vary with a parameter $n$
according to $p_l=n(1-p_s)\equiv p$. Such graphs then lie
along ``small-world" trajectories parameterized by $n$ and $p$ in
the space of graphs with general $(p_s, p_l)$. Previous solutions
of bond percolation on graphs with general $(p_s, p_l)$ have been
shown to give results in the same class as random 
graphs~\cite{MN1,MN2,KK}, with the transition between the percolating and
non-percolating phase controlled by mean-field exponents
$\alpha=-1$, $\beta=1$, and $\gamma=1$.

An interesting aspect of the the original Watts--Strogatz model,
and its generalization presented here is the interplay
between the deleted short-range links and the added shortcuts.
In this paper we have discussed several aspects of this interplay,
including the discontinuous jump between the two {\em percolating}
phases at $p=1$ and the critical behavior near the percolation
phase transition point.

We have presented an analytical solution of
generalized WS graphs, valid for $n \alt 1/4$, and  a
numerical solution for the standard ($n=1$) WS graph,
which together show that the critical behavior of such networks  in
the $p \rightarrow 0$ limit falls into one of three regimes of
behavior depending upon the value of $n$. Moreover, in none of
these regimes is the transition described by standard
mean-field. Trajectories for which $n>1/4$ display
a transition between two percolating phases in which the
percolation probability, $P$, jumps discontinuously from a finite
value to unity at the transition point while the mean
finite-cluster size, $S$, diverges. The critical exponents
associated with the transition are $\alpha=1$, $\beta=0$, and
$\gamma=1$. Trajectories with $n<1/4$ and $n=1/4$ both display a
percolation transition with a $P$ jumping discontinuously from 0
to 1 and critical exponents $\alpha=2$~\cite{n=0}
and $\beta=0$; however they
are differentiated by the behavior of $S$, which diverges with
exponent $\gamma=1$ for $n<1/4$ and $\gamma=2$ for $n=1/4$.

We have further proposed that all three regimes of behavior can be
unified within a single scaling form for the
cluster-size generating function, $G$, that is valid in the
vicinity of the critical point $p=0$. According to this proposal,
the failure of the critical exponents for $n\leq 1/4$ to satisfy
the identity $\alpha+2\beta+\gamma=2$ can be related to asymptotic
zeroes of the scaling function.
We have also considered the percolative behavior of generalized
WS graphs on higher-dimensional
lattices. While no exact solution exists, general considerations
of the form of the phase diagram for systems with general $(p_s,
p_ls)$ indicate that there should not be any critical behavior
along small--world trajectories.

To conclude, we note that, as far as we are aware, studies of the
dependence of the graph diameter, $l$, on the long-range bond
probability $p$ in WS graphs have so far focused
exclusively on the $n=1$ case. However, our analysis of bond
percolation in generalized WS graphs has shown that
varying $n$ can have important consequences in
case of a 1-dimensional underlying lattice. In particular,
according to the form of our scaling function, in the $p
\rightarrow 0$ limit it is the quantity $(n-1/4)/p$ that is the
physically important parameter in percolation. An interesting
question therefore arises as to whether the graph diameter $l(p)$
has a corresponding dependence on $n$, and in particular whether
the value $n=1/4$ has a physical significance beyond the
percolation behavior presented here.

It should be noted that the original WS model has been
extended mainly to $k$-rings, rings where each node is connected
to its $k$ nearest neighbors on each side (for a total of
$2k$ nearest neighbors bonds per site).
While this model is also one dimensional, and does not percolate without
long-range bonds (unless $p_s=1$), the singular behavior as $p\to 0$
could well be different from the case $k=1$.
For a $k$-ring, a lower
bound on the asymptotic probability that a node does not belong to a
long ``super-node'' extending to its left behaves as $(1-p_s)^k$. Using
Eq.~(\ref{y_p}) with $p_s'$ replacing $p_s$, and with the approximation
$1-p_s'\sim(1-p_s)^k$, one obtains that for $k>1$, $y\to 0$ for $p_s\to 1$.
Therefore, the behavior near the fully connected ring state is analytic---
in contrast to the singly connected ring no jump discontinuity exists.

\section*{Acknowledgments}
M.~K. is supported by NSF grant  DMR-04-26677.

\appendix
\section{The Generating Function Approach}
In this Appendix we develop a different approach to
studying the percolation transition in small world
networks. We use the generating function formalism,
stressing the topological, rather than thermodynamic
structure of the network.
The percolation process can be perceived as consisting
of two stages. In the first stage short range links
are removed from the network with probability $1-p_s$.
The remaining network consists of isolated islands
of varying sizes.
We will term a group of nodes connected by a chain of
short-range links a ``super-node.'' This construction can
be viewed as a renormalization of the network, where each
super-node is a node in the new network, having weight
proportional to its original size.
In the second stage long-range links are added to the network,
and we study the connectivity between super-nodes through
these long-range links. The probability, $P(d)$,
of a node to belong to a super-node of size $d$ is the number
of possible consecutive sets of size $d$ including this node
(which is just $d$) times the probability of the $d-1$ links
between these $d$ nodes to be intact, and the two links
leading to nodes neighboring this super-node to be removed.
Therefore,
\begin{equation}
P(d)=d p_s^{d-1}(1-p_s)^2\;,
\end{equation}
and the generating function for the super-node size is
\begin{equation}
S(x)=\sum_{d=1}^\infty d p_s^{d-1}(1-p_s)^2
x^d=\frac{x (1-p_s)^2}{1-xp_s}\;.
\end{equation}

Consider now the distribution of the number
of long range links emanating from a super-node
of size $d$, since there are $p_lN$ edges having
$2p_lN$ ends (links) randomly distributed on the
ring, we expect a Poisson distribution of the links
with mean $2p_l$ per node, or $2d p_l$ per super-node.
Therefore the probability of having $k$ links from a
super-node is
\begin{equation}
P(k|d)=\frac{(2d p_l)^ke^{-2dp_l}}{k!}\;,
\end{equation}
and the joint generating function for super-node size
and degree is
\begin{eqnarray}
D(x,y)&=&\sum_{d,k}P(d)P(k|d)x^dy^k\nonumber\\
&=&(1-p_s)^2
\frac{xe^{2p_l(y-1)}}{(1-p_sxe^{2p_l(y-1)})^2}\;.
\end{eqnarray}

To calculate the distribution of component sizes obtained
by following a link, one should notice that the probability
of reaching each node by following a long range link is
the same. Therefore, the probability of reaching a
super-node is proportional to its size and similar to
the probability that a node belongs to a super-node of size $d$.
Hence, the distribution of cluster sizes reached by
following a link (which is similar to the distribution of
cluster sizes in general due to the Poisson distribution)
is given by the self consistent equation
\begin{equation}
\label{self_con}
y=D(x,y)\;.
\end{equation}
From this self-consistency condition all
cluster and percolation properties can be obtained.
In particular, when finding the solution of the equation
\begin{equation}
\label{y_p}
y_p=D(1,y_p),
\end{equation}
if $y_p=1$ there exists no giant component, while for
$y_p<1$ a giant component exists. The size of the giant
component is determined to be $S=1-y_p$.
As discussed before, the behavior as $p_s\to 1$, with
$p_l=1-p_s$ is special:
As this limit is approached,
$S\to \frac{\sqrt{3}}{2}\approx 0.866$, with a discontinuous
jump to $S=1$ for $p_s=1$ and $p_l=0$~!

Expanding Eq.~(\ref{self_con}) in powers of $\epsilon$
for $y=1-\epsilon$ and $x=1$ gives the
phase transition behavior for the model. One obtains,
\begin{equation}
1-\epsilon=1+\frac{2p_l(p_s+1)}{p_s-1}\epsilon+
\frac{4p_l^2(p_s^2+4p_s+1)}{2(p_s-1)^2}\epsilon^2+\ldots\;,
\end{equation}
leading to the critical percolation threshold at
\begin{equation}
\label{thresh}
\frac{2p_l(p_s+1)}{1-p_s}=1\;.
\end{equation}
Near the threshold  it can be seen that the quadratic term
in $\epsilon$ will always exist unless $p_s\to 1$
or $p_l\to 0$.
For a generalized $n$ the behavior depends on the expansion
of the linear terms in $\epsilon$ as a function of the
deviation from the critical point.
Furthermore, for the original WS
model with $p=p_l=1-p_s$ it is easily seen that the whole
$0\leq p\leq 1$ range is within the percolating regime.


\begin{thebibliography}{99}
\bibitem{note1} Readers comparing these exponents
to the ones in the paper by Moore and Newman~\cite{MN2} should note
their non-standard exponents $\sigma$ and $\tau$. In standard
percolation language, $\tau_{\text{MN}}=(\tau_{\text{perc}} -1)$
and $\sigma_{\text{MN}}=1/\gamma_{\text{perc}}$. For a full
treatment of the definitions of critical exponents in percolation,
see Ref.~\cite{SA}.
\bibitem{n=0} Readers familiar with one-dimensional percolation will be
surprised to see us claim that $\alpha=2$ for the $n=0$ case,
rather than the accepted result $\alpha=1$. The discrepancy arises
because the term in $G$ that is linear in $p$ is independent of
$\delta_n$ and $\bar{m}$, and so has been excluded from our
definition of $G^{\text{sing}}$. Another point to bear in mind is
that our formulation of the model is ill-suited to dealing with
$n=0$ due to the numerous $1/n$ terms in Eq.~(\ref{Gg}) that arise
from our definition of $n$. If instead we simply set $p_l=0$ in
Eq.~(\ref{G}), no such problems arise.
\end{thebibliography}
\end{document}